\documentstyle [aps,prb]{revtex}
\begin{document}

\title{Computational implementation of the Kubo formula for the static
conductance: application to two-dimensional quantum dots}

\author{J. A. Verg\'es}
\address{Instituto de Ciencia de Materiales de Madrid,
Consejo Superior de Investigaciones Cient\'{\i}ficas,\\
Cantoblanco, E-28049 Madrid, Spain}

\date{\today}

\maketitle

\begin{abstract}
Kubo formula is used to get the d.c conductance of a statistical
ensemble of two-dimensional clusters of the square lattice
in the presence of standard diagonal disorder,
a uniform magnetic field and random magnetic fluxes.
Working within a one-band tight-binding approach the calculation
is quite general.
The shape of the cluster is rectangular with ideal leads attached
to opposite corners. Both geometrical characteristics and physical
parameters can be easily selected.
The output is just the conductance of a system of given parameters or a
statistical ensemble of conductances measured for different
disorder realizations.

\noindent
Section 7.9 of the Program Library Index
\end{abstract}


\part*{PROGRAM SUMMARY}

\vspace{0.3 cm}
\noindent
{\it Title of program:} KUBO

\vspace{0.3 cm}
\noindent
{\it Catalogue identifier:}

\vspace{0.3 cm}
\noindent
{\it Distribution format:} Uuencoded gzipped tar file

\vspace{0.3 cm}
\noindent
{\it Computer for which the program is designed and others on which it
has been tested:} Alphaserver 1200, DECstation 3000 Model 400

\vspace{0.3 cm}
\noindent
{\it Computers:}

\vspace{0.3 cm}
\noindent
{\it Installations:}

\vspace{0.3 cm}
\noindent
{\it Operating systems or monitors under which the program has been tested:}
Digital Unix 4.0D

\vspace{0.3 cm}
\noindent
{\it Programming language used:} Fortran

\vspace{0.3 cm}
\noindent
{\it Memory required to execute with typical data:} 7.04 Mwords

\vspace{0.3 cm}
\noindent
{\it Number of bits in a word:} 64

\vspace{0.3 cm}
\noindent
{\it Number of processors used:} 1

\vspace{0.3 cm}
\noindent
{\it Has the code been vectorised or parallelized?:}
Vectorised version of inversion LAPACK subroutines are available
on the computer (DXML library)

\vspace{0.3 cm}
\noindent
{\it Number of bytes in distributed program, including test data, etc.:}
7305

\vspace{0.3 cm}

\vspace{0.3 cm}
\noindent
{\it Nature of the physical problem}

\noindent
Conductance evaluation plays a central role in the study of the physical
properties of quantum dots [1]. The finite mesoscopic system under study is
connected to infinite leads defining an infinitesimal voltage drop.
Kubo formula is used for the conductance calculation once the Green function
of the whole system is known. Dependence of the conductance on
the number of conducting
modes in the leads, size and shape of the dot, Fermi energy,
presence of conventional diagonal disorder or random
magnetic fluxes through the plaquettes and a uniform magnetic field
through the sample can be considered as needed.

\vspace{0.3 cm}
\noindent
{\it Method of solution}

\noindent
The calculation of the Green function of the system formed by the quantum
dots and the leads is the main concern of the program [2].
Firstly, the selfenergy due to a semiinfinite ideal lead is calculated.
Secondly, this selfenergy is used as the starting value for an iteration
through the sample that takes into account its shape and local
disorder and magnetic fluxes.
Thirdly, the final selfenergy is matched via a two-slab Green function
calculation with the selfenergy coming from the opposite sample side.
Finally, Kubo formula is used for obtaining the system conductance
from the current traversing the dot.

\vspace{0.3 cm}
\noindent
{\it Restrictions on the complexity of the program}

\noindent
The more important restriction is the use of just a one-band tight-binding
Hamiltonian. Other program characteristics, like the rectangular shape of
the sample, the position of the leads near opposite sample corners, etc.
can be easily changed.

\vspace{0.3 cm}
\noindent
{\it Typical running time}

\noindent
The calculation of the conductance at a fixed Fermi energy for ten random
realizations of $100 \times 100$ disordered samples takes 225 seconds
on a DECstation 3000 Model 400.

\vspace{0.3 cm}
\noindent
{\it References}

\noindent [1]
For reviews of mesoscopic physics see
C.W.J. Beenakker and H. van Houten, in {\it Solid State Physics},
edited by H. Ehrenreich and D. Turnbull (Academic Press, New
York, 1991), Vol. 44, pp. 1--228;
{\it Mesoscopic Phenomena in Solids}, edited by B.L. Altshuler,
P.A. Lee, and R.A Webb (North--Holland, New York, 1991).

\noindent [2]
The following book by Datta gives a good account of both the basic concepts
and the practical things that should be solved to get the conductance
of a mesoscopic system,
S. Datta, {\it Electronic Transport in Mesoscopic Systems},
(Cambridge University Press, Cambridge, 1995).

\part*{LONG WRITE-UP}

\section{Introduction}

This paper deals with the numerical calculation of the conductance
of a two-dimensional system via Kubo formula \cite{kubo}
which is the standard microscopic approach to the problem.
It has been proved that it is equivalent to the more intuitive
point of view of expressing the conductance as a sum of transmission
coefficients over channels \cite{equivalence}.
Actually, a transmittance calculation for a disordered square lattice has been
implemented as an example of the use of the block recursion method
\cite{haydock}.

Both measurement and calculation of the conductance of mesoscopic
systems is an important issue in the study of their physical
properties \cite{mesoscopic}.
The calculation is as realistic as it is possible within
present computer facilities. Actually, the main limitation of the program
is just the consideration of a one-band tight-binding Hamiltonian.
In principle, this one-orbital per site basis can be enlarged within the
present scheme. The necessary effort makes only sense
when simulating a particular physical system.
On the other hand, the one-band tight-binding
basis can also be seen as the first approach to free electrons
moving in two-dimensions. Certainly, the lower part of the band shows
a quadratic dispersion relation.

Following well-known formalisms \cite{verges,datta}, the system includes
both a transversal uniform magnetic field and a random
component giving variable fluxes through the plaquettes.
Conventional diagonal disorder is also considered for completeness.
Further possibilities like other kinds of non-diagonal disorder are
absent but would be easily incorporated if necessary.
Also size of the sample, shape, number of modes transported by the leads
can be tuned at convenience. Again further complications like the
inclusion of more leads of different widths or going to a tunneling
regime in the contacts between sample and leads
only imply minor changes in the computer code.
In order to keep the program understandable,
I have left such kind of modifications for the readers.
Another interesting generalization is the calculation of three-dimensional
conductances. This has been actually done in the study of manganites
\cite{manganitas}.

\section{Lattice model}

The system is composed by a left ideal lead allowing up to $m$ modes,
an $L \times M$ cluster and a right ideal lead similar to the left one.
Leads are semiinfinite and are connected to opposite corners of the
rectangular dot by ideal contacts.
Sites form a subspace of the whole square lattice.
The Hamiltonian describing the quantum system is:
\begin{equation}
{\widehat H}_0 = \sum_i \sum_{j=1}^{M_i}
\epsilon_{i,j} {\hat c}_{i,j} \,\! ^\dag {\hat c}_{i,j} - t
\sum_i \sum_{j=1}^{M_i-1} \left(
e^{i 2 \pi \phi_{i,j}^{i+1,j}} {\hat c}_{i,j} \,\! ^\dag {\hat c}_{i+1,j} +
{\hat c}_{i,j} \,\! ^\dag {\hat c}_{i,j+1} +
e^{-i 2 \pi \phi_{i,j}^{i+1,j}} {\hat c}_{i+1,j} \,\! ^\dag {\hat c}_{i,j} +
{\hat c}_{i,j+1} \,\! ^\dag {\hat c}_{i,j}
\right) \qquad ,
\label{RMF}
\end{equation}
where ${\hat c}_{i,j} \,\! ^\dag$ creates an electron on site $(i,j)$,
$\epsilon_{i,j}$ are random diagonal energies chosen with
equal probability from the interval $[-W/2,W/2]$,
and $- t$ is the hopping energy.
Sums in $j$ depend on the local width $M_i$ which is $m$ for the leads
and $M$ for the dot. The flux through the lattice loop
which left lower corner is ${i,j}$ is given by:
$\Phi_{i,j} = \phi_{i,j+1}^{i+1,j+1} - \phi_{i,j}^{i+1,j}$
measured in units of flux quantum $\Phi_0 = {{hc} / e}$.
Uncorrelated fluxes are randomly selected with equal probability
from the interval
$[- {\phi_{max} / (2 \pi)},{\phi_{max} / (2 \pi)}]$.
A transversal uniform magnetic field $h_\perp$ is easily incorporated
adding to all $\Phi_{i,j}$ within the sample
the number of quantum fluxes traversing any plaquette
of the dot (${h_\perp a^2} / \Phi_0$, $a$ being the lattice constant).
With this gauge choice, when the system is solved layer by layer,
the matrix giving intralayer hopping is constant while the interlayer
coupling includes all effects due to the magnetic field.
Both disorder and magnetic fluxes are restricted to the dot, that is,
have non-zero values for $1 \le i \le L$.
It is also implicitly assumed that the number of modes (width) of the leads
$m$ is smaller than or equal to the dot width $M$. Nevertheless,
a constriction (i.e., the $m > M$ case) can be similarly handled.

\section{d.c. conductance via Kubo formula}
In order to study the linear electric response of the system, an electric
potential profile should be assumed. The simplest possibility is just
a stepwise potential: all dot and right lead sites ($i \ge 1$)
are shift by an electric potential energy equal to $-e V$ whereas left
lead sites ($i < 1$) remain at 0. This choice defines a position operator:
\begin{equation}
{\hat x} = \sum_{i \ge 1} \sum_{j=1}^{M_i}
{\hat c}_{i,j} \,\! ^\dag {\hat c}_{i,j}
\label{x}
\end{equation}
and the corresponding perturbing Hamiltonian:
\begin{equation}
{\widehat H}_1 = - e V \cos (\omega t) \hat x \qquad ,
\label{H1}
\end{equation}
where a simple cosine time dependence is included.
$\widehat H_1$ produces transitions among the stationary eigenstates
of the system and a time dependent current develops.
If the charge at the right part of the system is defined by:
\begin{equation}
{\hat Q} = - e \sum_{i \ge 1} \sum_{j=1}^{M_i}
{\hat c}_{i,j} \,\! ^\dag {\hat c}_{i,j}
= - e {\hat x} \qquad ,
\label{Q}
\end{equation}
the charge traversing the sample per unit time, i.e., the current through
the system is given by operator $\hat I$ that obeys the following
equation of motion:
\begin{equation}
i \hbar {\hat I} = \left [ {\widehat H},{\hat Q} \right ]
 = - e \left [ {\widehat H},{\hat x} \right ] = - e i \hbar {\hat v}_x
\qquad .
\label{I}
\end{equation}
Therefore, the current operator is the velocity operator apart from a
constant:
$\hat I = -e {\hat v}_x$.
Using the total Hamiltonian of the system
$\widehat H ={\widehat H}_0+{\widehat H}_1$
and the equation of motion for $\hat x$,
the first important result is obtained:
\begin{equation}
i \hbar {\hat v}_x = - t \sum_{j=1}^m \left(
{\hat c}_{0,j} \,\! ^\dag {\hat c}_{1,j} - 
{\hat c}_{1,j} \,\! ^\dag {\hat c}_{0,j}
\right) \qquad .
\label{vx}
\end{equation}
This equation shows that the knowledge of the charge flux in the contact
between the left lead and the sample is enough to evaluate the total
current through the system. Of course, the result comes from the flux
conservation along the whole system.

Now, first order time-dependent perturbation theory is used to get
the time evolution of the stationary eigenfunctions\cite{kubo}.
The mean value of the current is:
\begin{equation}
<\hat I>=\sum_\alpha f_\alpha
<\Psi_\alpha|\hat I|\Psi_\alpha> \qquad,
\label{current}
\end{equation}
where $\Psi_\alpha$ are the perturbed eigenfunctions and
$f_\alpha$ their occupation.
A straightforward but tedious calculation gives the linear electric
response of the system:
\begin{equation}
G=\lim_{\omega \rightarrow 0} \frac{<\hat I>}{V}=
-e^2 \hbar \pi \lim_{\omega \rightarrow 0}
\sum_{\alpha,\beta}{
{|<\alpha|{\hat v}_x|\beta>|}^2 \,
\frac{f_\alpha-f_\beta}{\epsilon_\alpha-\epsilon_\beta} \,
\delta(\epsilon_\beta-\epsilon_\alpha-\hbar \omega)
} \qquad,
\label{conductance}
\end{equation}
being $G$ the system conductance. To arrive at this formula, the
basic relation between the matrix elements of position and velocity
operators implied by the equation of motion (\ref{I}) has been used:
\begin{equation}
i \hbar <\alpha|{{\hat v}_x}|\beta>=
(\epsilon_\alpha - \epsilon_\beta) <\alpha|{\hat x}|\beta> \qquad.
\label{conmutador}
\end{equation}
Former conductivity calculations based on Kubo formalism used
Eq. \ref{conductance} to calculate the static conductivity\cite{stein_krey}.
Nevertheless, direct use of Eq. \ref{conductance}
leads to a lot of numerical troubles since eigenvalues are discrete for
isolated samples and the conductance is obtained as a sum of delta functions.
This fact forces the use of some averaging
procedure over finite $\omega$ values that distorts the final static
($\omega=0$) value.

When Green functions are used instead of eigenfunctions to express
the conductance, a more convenient form of the Kubo formula at
$T=0$ is obtained:
\begin{equation}
G = 2 {{\left( \frac{e^2}{h} \right)} {{\rm Tr} \left [(i \hbar {\hat v}_x)
{\rm Im\,}{\mathcal \widehat G}(E)(i \hbar {\hat v}_x)
{\rm Im\,}{\mathcal \widehat G}(E)\right ]}} \;,
\label{kubo}
\end{equation}
where ${\rm Im\,}{\mathcal \widehat G}(E)$ is obtained from
the advanced and retarded Green functions:
\begin{equation}
{\rm Im\,}{\mathcal \widehat G}(E)=\frac{1}{2i}\left[{\mathcal
\widehat G}^{R}(E)-{\mathcal \widehat G}^{A}(E)\right ] \;,
\label{img}
\end{equation}
and the energy $E$ is the Fermi energy of the system.
Traditionally, the static electrical conductivity $\sigma_{xx}$ was
calculated\cite{nozieres}. It coincides with $G$ in a two-dimensional
system. Conductance is given in units of $e^2/h$
which is the quantum conductance unit. The trace in
Eq. \ref{kubo} is dimensionless since the energy dimension of
$i \hbar {\hat v}_x$ is cancelled by the inverse energy dimension of the
imaginary part of the Green function.

Eq.\ref{kubo} shows that Green functions is the last ingredient needed
for the calculation of the conductance. Actually, only Green function matrix
elements at slabs $i=0$ and $i=1$ are necessary owing to the simple form
of ${\hat v}_x$ (see Eq. \ref{vx}) and the presence of the trace.
This feature dictates the method for choice in the Green function calculation.
Starting from the right end,
the selfenergy of the ideal right lead is obtained,
then iterated through the sample layer by layer, and finally, connected
to the selfenergy coming from the left ideal lead (which in fact coincides
with the right lead selfenergy).
The equation defining the iteration is obtained from Dyson equation for
the Green function\cite{thouless} and has the following form:
\begin{equation}
{\bf \Sigma}_{i-1}={\bf V}_{i-1,i}
(E {\bf I} - {\bf V}_{i,i} - {\bf \Sigma}_i)^{-1}
{\bf V}_{i,i-1} \qquad.
\label{iteracion}
\end{equation}
This allows the calculation of ${\bf \Sigma}_{i-1}$ once the former
selfenergy matrix and the intralayer ${\bf V}_{i,i}$ and interlayer
${\bf V}_{i-1,i}$ (${\bf V}_{i,i-1} = {\bf V}_{i-1,i} \,\! ^\dag$)
matrices are known.
After iterating the right side from $+\infty$ to 1 and the left side
from $-\infty$ to 0, the needed Green function matrix elements are given by:
\begin{equation}
[E {\bf I} - {\bf H} - {\bf \Sigma}_{\mathrm l} (E) -
{\bf \Sigma}_{\mathrm r}(E)] {\bf {\mathcal G}}(E) = {\bf I} \qquad,
\label{green}
\end{equation}
where $\bf H$ is the matrix representing the layer 0 plus layer 1
system Hamiltonian and ${\bf \Sigma}_{\mathrm l(r)}(E)$ stand for
the left (right) selfenergy matrices.

Alternatively, instead of iterating the ordered ideal leads, one can
analytically solve the problem in the basis of the transversal
modes\cite{datta}.
Later on, a basis transformation provides the desired selfenergy matrices.
Specifically, the retarded selfenergy due to the mode of wavevector
$k_y$ at energy $E$ is given by:
\begin{equation}
\Sigma(k_y)={1 \over 2}
\left( E  -\varepsilon (k_y)
-i\sqrt{4t^2-(E - \varepsilon (k_y))^2} \right) \; ,
\label{sigma1}
\end{equation}
within its band ($|E - \varepsilon (k_y)| \le 2 |t|$) and by:
\begin{equation}
\Sigma(k_y)={1 \over 2}
\left( E - \varepsilon (k_y) \mp
\sqrt{(E -\varepsilon (k_y))^2-4t^2} \right) \; ,
\label{sigma2}
\end{equation}
outside the band (minus sign for $E > \varepsilon (k_y)+2 |t|$
and plus sign for $E < \varepsilon (k_y))-2 |t|$), where
$\varepsilon (k_y)=2 t \cos (k_y a)$ is the
eigenenergy of the $k_y$ mode which is quantized:
$$
k_y={{i_y \pi} \over {(m+1)a}}   ,
$$
$i_y$ being an integer from 1 to $m$.
The transformation from normal modes to the local tight-binding basis
is obtained from the amplitudes of the normal modes:
\begin{equation}
<n_y|k_y>=\sqrt{2 \over {m +1}} \sin(k_y n_y a) \; ,
\label{eigenvectors}
\end{equation}
where $n_y$ is the tight-binding orbital position.
The elements of the retarded selfenergy matrix are thus: 
\begin{equation}
<n_{y_1}|{\widehat \Sigma} (E)|n_{y_2}>=
\sum_{k_y} {<n_{y_1}|k_y> \Sigma(k_y) <n_{y_2}|k_y>}
\label{sigmamatrix}
\end{equation}
Advanced Green function matrix elements are obtained as the transposed
complex conjugate elements of the retarded Green function matrix elements.

Kubo formula (Eq.\ref{kubo}) shows that since $\hat x$ operator is chosen
dimensionless,
the hopping integral $t$ appears twice in the numerator and also twice in the
denominator of the trace. Consequently, its value is irrelevant in the
conductance evaluation and can be set equal to 1 within the computer code
for simplicity. Apart from the iterative calculation of lead selfenergy
and a bit more restrictive geometrical set-up, the paper by Lee and
Fisher\cite{first} seems to be the first application of the formalism
presented in the last two sections.

\section{Program structure}

The structure of the program is extremely simple. Subroutines are not
necessary. Hard work is done by {\it state of the art} inversion
subroutines. An enumeration of the main program steps followed by a simple
explanation is enough for its understanding.

After reading the input and logging it for further reference,
both the intralayer matrix interaction ${\bf V}_{i,i}$ (which does not
depend on $i$ owing to the gauge choice) and the lead
eigenvectors (Eq. \ref{eigenvectors}) are initialized.
For a given Fermi energy $E$, selfenergies are calculated in the
eigenvectors basis using Eqs. \ref{sigma1} or \ref{sigma2}
and transformed to the tight-binding
basis using Eq. \ref{sigmamatrix}. The resulting selfenergy matrix is
called $tt$ within the program. Here, we arrived at the main part of the
calculation: selfenergy matrix is iterated from right to left
through a randomly generated
sample that incorporates the desired disorder. Once the selfenergy is known
at the left side of the sample, it is matched to the lead selfenergy
coming from the left by Eq.\ref{green}.
This allows the evaluation of the imaginary part of the Green
function via Eq. \ref{img} and the use of Kubo formula for the obtaining of
the conductance (see Eq. \ref{kubo}) that is the only program output
(Conductance is written in units of the quantum unit $e^2 / h$).
Notice that the multiplication by the matrix representing the
velocity operator in the local tight-binding basis (Eq. \ref{vx})
is done without building explicitly the very sparse matrix.
The calculation of the conductance
is repeated the number of times that are necessary for the
knowledge of the statistical behavior of the conductance due to disorder.
Of course, the step is not repeated when disorder is absent.

The heavy part of the computational work is just the inversion of a large
number of matrices of moderate size. It is achieved by external calls
to LAPACK subroutines that are conveniently adapted to the hardware
configuration. If not belonging to the
machine software, they are available via Internet (http://www.netlib.org).

A portable simple version of a generator of pseudo-random numbers is included
within the distributed source. This allows to check the distributed sample
results. Nevertheless, a better generator should be used
in the working version of the program. Notice that the generator included
within the source code
has a maximum period of 259200 which is really insufficient when a good
disorder statistics of large samples should be obtained. In any case, let
me remind that the seed of the generator should be smaller than $im$, i.e.,
smaller than 259200.

\section{Input and output}

Input parameters have all a simple physical meaning. They are given
in Table I. Output is just the conductance as a function of the
parameters (Fermi energy, for example) that are varied within the
external do loops of the program.
Conductance is given in units of its quantum $e^2/h$.

\section{Test run}

Input and output files corresponding to three typical calculations
are included in the distribution.
The first one gives the conductance of a small cluster in the presence
of a uniform magnetic field. It is obtained as a function of the Fermi
energy. This run does not use the pseudo-random numbers generator.
The second one studies the effect of disorder. One hundred samples
are randomly generated and their conductance calculated.
The third one shows the effect of geometric parameters. While the two
previous runs deal with a highly symmetric system ($32 \times 32$ cluster
attached to leads of the same width), this case gives the conductance
for a less symmetric dot.

One additional test is also possible. Conductance is an integer when
disorder is absent and leads perfectly match opposite rectangle sides.
It gives just the number of open (conducting) channels (modes) in the leads.

The same fortran code has been compiled and run on a Pentium II in a Linux
environment using g77 and local inversion subroutines. Results do not
differ from the given output samples.

\acknowledgments
I acknowledge Luis Mart\'{\i}n-Moreno who helped me checking the computer
code used in this work running his own program based on a different
implementation of the transfer matrix technique.
I also thank Rafael Ramirez for checking the code on a PC running under
Linux. This work has been partially supported by Spanish Comisi\'on
Interministerial de Ciencia y Tecnolog\'{\i}a (grant PB96-0085).

\part*{TEST RUN INPUT}

\subsection{Input 1}
\begin{verbatim}
032                       width
032                       length
032                       modes
0.0d0                     phimax
0.101341699d0             hmag
1751                      loteria
1                         ntimes
0.0d0 0.0d0 1.0d0         W1,W2,W3
-4.0d0 -3.0d0 0.01d0      energy1,energy2,energy3
\end{verbatim}

\subsection{Input 2}
\begin{verbatim}
032                       width
032                       length
032                       modes
3.14159265d0              phimax
0.0d0                     hmag
1751                      loteria
100                       ntimes
0.1d0 0.1d0 1.0d0         W1,W2,W3
0.1751d0 1.0d0 1.0d0      energy1,energy2,energy3
\end{verbatim}

\subsection{Input 3}
\begin{verbatim}
028                       width
038                       length
006                       modes
0.0d0                     phimax
0.0d0                     hmag
1751                      loteria
1                         ntimes
0.0d0 0.0d0 1.0d0         W1,W2,W3
-4.0d0 -3.0d0 0.01d0      energy1,energy2,energy3
\end{verbatim}

\part*{TEST RUN OUTPUT}

\subsection{Output 1}
\begin{verbatim}
 sample disorder_width  Fermi_level      conductance

   1        0.0000       -4.000000      0.52279883E-30
   1        0.0000       -3.990000      0.37576695E-15
   1        0.0000       -3.980000      0.10174681E-13
   1        0.0000       -3.970000      0.35160914E-13
   1        0.0000       -3.960000      0.13971441E-11
   1        0.0000       -3.950000      0.90081246E-11
   1        0.0000       -3.940000      0.19577031E-10
   1        0.0000       -3.930000      0.26700263E-09
   1        0.0000       -3.920000      0.73710663E-09
   1        0.0000       -3.910000      0.58263154E-07
   1        0.0000       -3.900000      0.15099144E-02
   1        0.0000       -3.890000      0.99895187E+00
   1        0.0000       -3.880000      0.99605711E+00
   1        0.0000       -3.870000      0.98332107E+00
   1        0.0000       -3.860000      0.99814562E+00
   1        0.0000       -3.850000      0.99428901E+00
   1        0.0000       -3.840000      0.99899589E+00
   1        0.0000       -3.830000      0.99661623E+00
   1        0.0000       -3.820000      0.99941302E+00
   1        0.0000       -3.810000      0.99948560E+00
   1        0.0000       -3.800000      0.99773298E+00
   1        0.0000       -3.790000      0.99865209E+00
   1        0.0000       -3.780000      0.99974332E+00
   1        0.0000       -3.770000      0.99753821E+00
   1        0.0000       -3.760000      0.99946904E+00
   1        0.0000       -3.750000      0.99999165E+00
   1        0.0000       -3.740000      0.99962089E+00
   1        0.0000       -3.730000      0.99937579E+00
   1        0.0000       -3.720000      0.99952032E+00
   1        0.0000       -3.710000      0.99980170E+00
   1        0.0000       -3.700000      0.19550936E+01
   1        0.0000       -3.690000      0.19914236E+01
   1        0.0000       -3.680000      0.19900318E+01
   1        0.0000       -3.670000      0.19861029E+01
   1        0.0000       -3.660000      0.19941228E+01
   1        0.0000       -3.650000      0.19974424E+01
   1        0.0000       -3.640000      0.19962072E+01
   1        0.0000       -3.630000      0.19979441E+01
   1        0.0000       -3.620000      0.19983424E+01
   1        0.0000       -3.610000      0.19973790E+01
   1        0.0000       -3.600000      0.19977287E+01
   1        0.0000       -3.590000      0.19987828E+01
   1        0.0000       -3.580000      0.19972511E+01
   1        0.0000       -3.570000      0.19933848E+01
   1        0.0000       -3.560000      0.19989170E+01
   1        0.0000       -3.550000      0.19995162E+01
   1        0.0000       -3.540000      0.19988242E+01
   1        0.0000       -3.530000      0.19985759E+01
   1        0.0000       -3.520000      0.19989692E+01
   1        0.0000       -3.510000      0.20108456E+01
   1        0.0000       -3.500000      0.29588509E+01
   1        0.0000       -3.490000      0.29880224E+01
   1        0.0000       -3.480000      0.29890605E+01
   1        0.0000       -3.470000      0.29934051E+01
   1        0.0000       -3.460000      0.29829712E+01
   1        0.0000       -3.450000      0.29908045E+01
   1        0.0000       -3.440000      0.29939537E+01
   1        0.0000       -3.430000      0.29913380E+01
   1        0.0000       -3.420000      0.29956061E+01
   1        0.0000       -3.410000      0.29991560E+01
   1        0.0000       -3.400000      0.29996230E+01
   1        0.0000       -3.390000      0.29985708E+01
   1        0.0000       -3.380000      0.29977878E+01
   1        0.0000       -3.370000      0.29978828E+01
   1        0.0000       -3.360000      0.29985529E+01
   1        0.0000       -3.350000      0.29991855E+01
   1        0.0000       -3.340000      0.29992141E+01
   1        0.0000       -3.330000      0.29981113E+01
   1        0.0000       -3.320000      0.39730935E+01
   1        0.0000       -3.310000      0.39118845E+01
   1        0.0000       -3.300000      0.39902136E+01
   1        0.0000       -3.290000      0.39906815E+01
   1        0.0000       -3.280000      0.39881601E+01
   1        0.0000       -3.270000      0.39945268E+01
   1        0.0000       -3.260000      0.39981701E+01
   1        0.0000       -3.250000      0.39966727E+01
   1        0.0000       -3.240000      0.39951492E+01
   1        0.0000       -3.230000      0.39961325E+01
   1        0.0000       -3.220000      0.39978480E+01
   1        0.0000       -3.210000      0.39980835E+01
   1        0.0000       -3.200000      0.39968286E+01
   1        0.0000       -3.190000      0.39955763E+01
   1        0.0000       -3.180000      0.39952847E+01
   1        0.0000       -3.170000      0.39950792E+01
   1        0.0000       -3.160000      0.39859467E+01
   1        0.0000       -3.150000      0.39946114E+01
   1        0.0000       -3.140000      0.40394524E+01
   1        0.0000       -3.130000      0.49848798E+01
   1        0.0000       -3.120000      0.49922641E+01
   1        0.0000       -3.110000      0.49917257E+01
   1        0.0000       -3.100000      0.49932667E+01
   1        0.0000       -3.090000      0.49965445E+01
   1        0.0000       -3.080000      0.49970599E+01
   1        0.0000       -3.070000      0.49964073E+01
   1        0.0000       -3.060000      0.49958574E+01
   1        0.0000       -3.050000      0.49957888E+01
   1        0.0000       -3.040000      0.49962471E+01
   1        0.0000       -3.030000      0.49966956E+01
   1        0.0000       -3.020000      0.49961112E+01
   1        0.0000       -3.010000      0.49934337E+01
   1        0.0000       -3.000000      0.49804470E+01


\end{verbatim}

\subsection{Output 2}
\begin{verbatim}
 sample disorder_width  Fermi_level      conductance

   1        0.1000        0.175100      0.15757571E+01
   2        0.1000        0.175100      0.90865074E+00
   3        0.1000        0.175100      0.89621507E+00
   4        0.1000        0.175100      0.17411444E+01
   5        0.1000        0.175100      0.17895352E+01
   6        0.1000        0.175100      0.16377963E+01
   7        0.1000        0.175100      0.13944614E+01
   8        0.1000        0.175100      0.17683471E+01
   9        0.1000        0.175100      0.14445051E+01
  10        0.1000        0.175100      0.15230297E+01
  11        0.1000        0.175100      0.14065148E+01
  12        0.1000        0.175100      0.15391601E+01
  13        0.1000        0.175100      0.14052089E+01
  14        0.1000        0.175100      0.13567799E+01
  15        0.1000        0.175100      0.16651201E+01
  16        0.1000        0.175100      0.17878714E+01
  17        0.1000        0.175100      0.11109133E+01
  18        0.1000        0.175100      0.89038712E+00
  19        0.1000        0.175100      0.12773366E+01
  20        0.1000        0.175100      0.11877100E+01
  21        0.1000        0.175100      0.79052484E+00
  22        0.1000        0.175100      0.14452282E+01
  23        0.1000        0.175100      0.19539141E+01
  24        0.1000        0.175100      0.14051744E+01
  25        0.1000        0.175100      0.13141107E+01
  26        0.1000        0.175100      0.15579377E+01
  27        0.1000        0.175100      0.17998757E+01
  28        0.1000        0.175100      0.15839267E+01
  29        0.1000        0.175100      0.15788136E+01
  30        0.1000        0.175100      0.18356223E+01
  31        0.1000        0.175100      0.16829076E+01
  32        0.1000        0.175100      0.16927771E+01
  33        0.1000        0.175100      0.14242584E+01
  34        0.1000        0.175100      0.20767889E+01
  35        0.1000        0.175100      0.13799149E+01
  36        0.1000        0.175100      0.13587508E+01
  37        0.1000        0.175100      0.96175170E+00
  38        0.1000        0.175100      0.10228039E+01
  39        0.1000        0.175100      0.12647448E+01
  40        0.1000        0.175100      0.15151721E+01
  41        0.1000        0.175100      0.14502308E+01
  42        0.1000        0.175100      0.15174573E+01
  43        0.1000        0.175100      0.12829172E+01
  44        0.1000        0.175100      0.19077400E+01
  45        0.1000        0.175100      0.18329162E+01
  46        0.1000        0.175100      0.14944505E+01
  47        0.1000        0.175100      0.15520437E+01
  48        0.1000        0.175100      0.15797294E+01
  49        0.1000        0.175100      0.15870156E+01
  50        0.1000        0.175100      0.19959013E+01
  51        0.1000        0.175100      0.19913650E+01
  52        0.1000        0.175100      0.12217520E+01
  53        0.1000        0.175100      0.16039270E+01
  54        0.1000        0.175100      0.17024037E+01
  55        0.1000        0.175100      0.14668939E+01
  56        0.1000        0.175100      0.15284083E+01
  57        0.1000        0.175100      0.12557186E+01
  58        0.1000        0.175100      0.12838045E+01
  59        0.1000        0.175100      0.18537063E+01
  60        0.1000        0.175100      0.16127581E+01
  61        0.1000        0.175100      0.15376409E+01
  62        0.1000        0.175100      0.15619712E+01
  63        0.1000        0.175100      0.17955325E+01
  64        0.1000        0.175100      0.14001172E+01
  65        0.1000        0.175100      0.97460353E+00
  66        0.1000        0.175100      0.11018902E+01
  67        0.1000        0.175100      0.15973680E+01
  68        0.1000        0.175100      0.13172499E+01
  69        0.1000        0.175100      0.89201764E+00
  70        0.1000        0.175100      0.12745779E+01
  71        0.1000        0.175100      0.20770076E+01
  72        0.1000        0.175100      0.12221830E+01
  73        0.1000        0.175100      0.16596201E+01
  74        0.1000        0.175100      0.96965211E+00
  75        0.1000        0.175100      0.13451393E+01
  76        0.1000        0.175100      0.15577793E+01
  77        0.1000        0.175100      0.12640805E+01
  78        0.1000        0.175100      0.20086567E+01
  79        0.1000        0.175100      0.14018443E+01
  80        0.1000        0.175100      0.14841935E+01
  81        0.1000        0.175100      0.11329059E+01
  82        0.1000        0.175100      0.10930836E+01
  83        0.1000        0.175100      0.17438574E+01
  84        0.1000        0.175100      0.11191919E+01
  85        0.1000        0.175100      0.16853219E+01
  86        0.1000        0.175100      0.13615257E+01
  87        0.1000        0.175100      0.12996388E+01
  88        0.1000        0.175100      0.14827696E+01
  89        0.1000        0.175100      0.15302789E+01
  90        0.1000        0.175100      0.17095241E+01
  91        0.1000        0.175100      0.11144200E+01
  92        0.1000        0.175100      0.15511103E+01
  93        0.1000        0.175100      0.17014254E+01
  94        0.1000        0.175100      0.13148788E+01
  95        0.1000        0.175100      0.12656503E+01
  96        0.1000        0.175100      0.12587740E+01
  97        0.1000        0.175100      0.11158110E+01
  98        0.1000        0.175100      0.16342194E+01
  99        0.1000        0.175100      0.17836377E+01
 100        0.1000        0.175100      0.15009786E+01


\end{verbatim}

\subsection{Output 3}
\begin{verbatim}
 sample disorder_width  Fermi_level      conductance

   1        0.0000       -4.000000      0.41985273E-31
   1        0.0000       -3.990000      0.71259408E-32
   1        0.0000       -3.980000      0.13096324E-31
   1        0.0000       -3.970000      0.89940928E-31
   1        0.0000       -3.960000      0.84981659E-32
   1        0.0000       -3.950000      0.74052006E-31
   1        0.0000       -3.940000      0.10438540E-30
   1        0.0000       -3.930000      0.56327809E-32
   1        0.0000       -3.920000      0.86474255E-31
   1        0.0000       -3.910000      0.21088933E-31
   1        0.0000       -3.900000      0.38364524E-30
   1        0.0000       -3.890000     -0.20299302E-30
   1        0.0000       -3.880000      0.26000054E-32
   1        0.0000       -3.870000      0.83486242E-31
   1        0.0000       -3.860000      0.30814879E-31
   1        0.0000       -3.850000      0.20360691E-30
   1        0.0000       -3.840000      0.31061398E-29
   1        0.0000       -3.830000      0.44989724E-30
   1        0.0000       -3.820000     -0.80580909E-30
   1        0.0000       -3.810000      0.14202891E-29
   1        0.0000       -3.800000      0.12988896E+00
   1        0.0000       -3.790000      0.43092998E-01
   1        0.0000       -3.780000      0.19481001E+00
   1        0.0000       -3.770000      0.22292412E+00
   1        0.0000       -3.760000      0.77581649E+00
   1        0.0000       -3.750000      0.77562524E+00
   1        0.0000       -3.740000      0.66989858E+00
   1        0.0000       -3.730000      0.28693757E-01
   1        0.0000       -3.720000      0.67556799E+00
   1        0.0000       -3.710000      0.98570186E+00
   1        0.0000       -3.700000      0.97495692E+00
   1        0.0000       -3.690000      0.99581430E+00
   1        0.0000       -3.680000      0.97023908E-01
   1        0.0000       -3.670000      0.82385509E+00
   1        0.0000       -3.660000      0.91641977E-02
   1        0.0000       -3.650000      0.48033012E-01
   1        0.0000       -3.640000      0.88955232E+00
   1        0.0000       -3.630000      0.85081542E+00
   1        0.0000       -3.620000      0.66479842E+00
   1        0.0000       -3.610000      0.64385114E+00
   1        0.0000       -3.600000      0.99855320E+00
   1        0.0000       -3.590000      0.28084687E+00
   1        0.0000       -3.580000      0.38374067E+00
   1        0.0000       -3.570000      0.10191610E+00
   1        0.0000       -3.560000      0.57763375E+00
   1        0.0000       -3.550000      0.96053453E+00
   1        0.0000       -3.540000      0.97642322E+00
   1        0.0000       -3.530000      0.98907542E+00
   1        0.0000       -3.520000      0.99980992E+00
   1        0.0000       -3.510000      0.99966610E+00
   1        0.0000       -3.500000      0.44537039E+00
   1        0.0000       -3.490000      0.16269503E+00
   1        0.0000       -3.480000      0.25320153E+00
   1        0.0000       -3.470000      0.33512914E+00
   1        0.0000       -3.460000      0.43581456E-01
   1        0.0000       -3.450000      0.86944911E+00
   1        0.0000       -3.440000      0.97491430E+00
   1        0.0000       -3.430000      0.77663803E+00
   1        0.0000       -3.420000      0.86652483E+00
   1        0.0000       -3.410000      0.17048486E+00
   1        0.0000       -3.400000      0.52572889E+00
   1        0.0000       -3.390000      0.99697782E+00
   1        0.0000       -3.380000      0.97017401E+00
   1        0.0000       -3.370000      0.76690068E-01
   1        0.0000       -3.360000      0.65180288E+00
   1        0.0000       -3.350000      0.91149142E-01
   1        0.0000       -3.340000      0.55435669E+00
   1        0.0000       -3.330000      0.78091884E+00
   1        0.0000       -3.320000      0.82541535E+00
   1        0.0000       -3.310000      0.79649338E+00
   1        0.0000       -3.300000      0.72273498E+00
   1        0.0000       -3.290000      0.80582563E+00
   1        0.0000       -3.280000      0.99113612E+00
   1        0.0000       -3.270000      0.99944946E+00
   1        0.0000       -3.260000      0.98089019E+00
   1        0.0000       -3.250000      0.34866254E+00
   1        0.0000       -3.240000      0.80427201E+00
   1        0.0000       -3.230000      0.10280973E+01
   1        0.0000       -3.220000      0.10774430E+01
   1        0.0000       -3.210000      0.99631329E+00
   1        0.0000       -3.200000      0.11419407E+01
   1        0.0000       -3.190000      0.78964694E+00
   1        0.0000       -3.180000      0.77838727E+00
   1        0.0000       -3.170000      0.12652009E+01
   1        0.0000       -3.160000      0.10966078E+01
   1        0.0000       -3.150000      0.12197723E+01
   1        0.0000       -3.140000      0.12790068E+01
   1        0.0000       -3.130000      0.99891612E+00
   1        0.0000       -3.120000      0.12284234E+01
   1        0.0000       -3.110000      0.10230287E+01
   1        0.0000       -3.100000      0.10131873E+01
   1        0.0000       -3.090000      0.14859354E+01
   1        0.0000       -3.080000      0.11383726E+01
   1        0.0000       -3.070000      0.10832137E+01
   1        0.0000       -3.060000      0.70474129E+00
   1        0.0000       -3.050000      0.83114497E+00
   1        0.0000       -3.040000      0.12279884E+01
   1        0.0000       -3.030000      0.11348148E+01
   1        0.0000       -3.020000      0.18572412E+01
   1        0.0000       -3.010000      0.15447665E+01
   1        0.0000       -3.000000      0.11718278E+01


\end{verbatim}

\newpage
\begin{table}
\caption{Physical meaning of the program parameters that are read on
unit 5}
\label{input}
\begin{tabular}{ll}
width & width of the sample, $M$ (integer*4)\\
length & length of the sample, $L$ (integer*4)\\
modes & number of modes in the leads, $m$ (integer*4)\\
phimax & half the width of the random flux distribution,
         $\phi_{max}$ (real*8)\\
hmag & uniform magnetic field in the sample, $h_\perp$ (real*8)\\
loteria & seed for the pseudorandom number generator (integer*4)\\
ntimes & number of samples in the statistical ensemble (integer*4)\\
W1 & initial width ($W$) of diagonal disorder distribution (real*8)\\
W2 & final width ($W$) of diagonal disorder distribution (real*8)\\
W3 & step in the disorder analysis (real*8)\\
energy1 & initial Fermi energy ($E$) (real*8)\\
energy2 & final Fermi energy ($E$) (real*8)\\
energy3 & step in the Fermi energy analysis (real*8)\\
\end{tabular}
\end{table}

\end{document}